# Version 6

# The cognitive homunculus: do tunable languages-of-thought convey adaptive advantage?

Rodrick Wallace, Ph.D.
The New York State Psychiatric Institute*

December 1, 2003

## Abstract

We reexamine the generalized cognitive homunculus, an organism's internalized image of its physiological, psychological, and social state, which, when properly adjusted, can quickly detect subtle deviations from a reference configuration. We particularly seek to extend the treatment beyond 'language-of-thought' systems modeled as ergodic information sources. Such extension would generate an exceedingly rich response repertoire, not limited by fixed patterns of grammar and syntax. Rather, these would themselves be tunable according to the changing short-term contextual demands faced by the organism, possibly providing significant long-term adaptive advantage.

### Introduction

Several recent papers (R. Wallace, 2003; R. Wallace and D. Wallace, 2003a, b) have used an information theory perspective to examine a generalized cognitive homunculus (GCH), the body's internal self-image of its physiological and psychosocial state which can efficiently detect deviations from a baseline 'normal' condition. The foundation of this work was a 'duality' between a certain class of nested cognitive processes and an appropriately parametized adiabatically, piecewise, memoryless ergodic information source. An essential point is that a cognitive system must both recognize a pattern that requires response and then choose one from a larger repertoire of possible responses. The resulting model generalizes Fodor's (1975, 1981, 1987, 1990, 1998, 1999) language-of-thought hypothesis which invokes a draconian Chomskian 'linguistic completeness' for internal cognitive workings.

The nested set of cognitive physiological, psychological, and social modules producing the GCH is characterized by a composite mutual information splitting criterion. This is defined through the embedding of the set of cognitive modules within a structured pattern of contextual stressors, itself having a well defined grammar and syntax, permitting identification as an information source. We refer to the previous development (Wallace et al., 2003a, b, c) for details.

Given an interacting set of processes, $Y_j, j = 1...s$, embedded in the larger context, $Z$, the appropriate mutual information is

$$I(Y_1,...,Y_s|Z) = H(Z) + \sum_{j=1}^{s} H(Y_j|Z) - H(Y_1,...,Y_s,Z)$$

where the $H$ are the respective individual, conditional, and joint, Shannon source uncertainties (e.g. Ash, 1990; Cover and Thomas, 1991).

Analysis suggests that a properly tuned GCH can detect exceedingly subtle changes in the condition of the body and its social relations, using a kind of retina argument, a generalization of the learning paradigms of conventional neural theory (e.g. Deco and Obradovic, 1996). Information theory treatments are notorious, however, for their chronic inability to specify an engineering strategy. Here we reexamine the 'thermodynamics' of a nest of cognitive modules, seeking to move beyond cognitive processes which must be modeled by ergodic information sources, i.e. going beyond languages-of-thought strongly constrained to a particular grammar and syntax.

As before, we take our underlying models from recent studies of punctuated evolutionary process on socially structured systems (Wallace and Wallace, 1998, 1999; Wallace et al., 2003a, b, c) and from Ives' (1995) study of linked ecosystem components, rather than from simple physical analogs based on the response of rate functions to entropy gradients (Onsager and Machlup, 1953; Kadtke and Bulsara, 1997). This leads toward a generalization of earlier work to 'non-ergodic' systems.

### The 'equation of state' of the GCH

Earlier work in this direction (e.g. R. Wallace, 2000; Wallace et al., 2003) focused on the properties of a cognitive system near a critical point, where the renormalization properties reflected the underlying 'neural architecture', in a large sense. We reexamine how such a system behaves far from a critical point.

Externally supplied 'sensory activity' and an internal 'ongoing activity', both of which we allow to be quite structured, are convoluted in a possibly complicated manner to produce a 'path' of signals $x = a_0, a_1, ..., a_n$ which is then fed into a 'nonlinear oscillator' having an output $h(x)$. We are particularly interested in paths beginning at some fixed $a_0$ for which $h(a_0, a_1, ..., a_m) \in B_0$ for all $m < n$ but for which $h(a_0, ..., a_n) \in B_1$, where $B_0$ and $B_1$ are sets of possible system responses. These paths we call, in the usual information theory terminology (e.g. Ash, 1990), 'meaningful.' Only a small number of all possible paths is assumed to be 'meaningful,' triggering $h$. Thus we have divided the possible outputs of the system, the $h(x)$, into two sets, $B_0$ and $B_1$, and look for paths whose outputs are initially all in $B_0$, and which then end in an element of $B_1$, the response chosen from the set of all possible responses. Note that division of the set of all possible responses into $B_0$ and $B_1$ may, itself, involve a 'higher order' cognitive process, a matter we will not further explore here but which leads to deep questions regarding autocognitive developmental disorders (R. Wallace, 2003).

---

*Address correspondence to Rodrick Wallace, PISCS Inc., 549 W. 123 St., Suite 16F, New York, NY, 10027. Tel. (212) 865-4766, rdwall@ix.netcom.com. Affiliation is for identification only.

It is precisely the disjunction of paths into a small number of high probability 'meaningful' states and a much larger number of others with very low probability that permits application of the Shannon-McMillan Theorem. This theorem, also called the Asymptotic Equipartition Theorem, states that the number $N(n)$ of meaningful paths of length $n$ satisfies the subadditive relation

$$H[\mathbf{X}] = \lim_{n \to \infty} \frac{\log[N(n)]}{n}$$

(1)

where $H[\mathbf{X}]$ is the source uncertainty defined from the joint and conditional probabilities of the paths $x$.

If the information source is stationary and ergodic, then $H[\mathbf{X}]$ is independent of path, a remarkable result. The probability of a meaningful path of length $n$, for large $n$, is thus $\propto \exp(nH[\mathbf{X}])$.

We will later attempt a generalization to 'slightly-less-than-ergodic' systems having (not too much) path dependence, taking us beyond the language-of-thought hypothesis.

See R. Wallace (2000) or Wallace et al. (2003a, b, c) for mathematical details. Indeed, this kind of relation serves as the starting point for most modern treatments of large deviations (e.g. Dembo and Zeitouni, 1998).

The definition can be extended, under proper conditions, to a stochastic system in which $h(x)$ is the probability that the nonlinear oscillator fires, provided a disjunction can still be made between paths $x$ which have high and low probabilities of triggering the oscillator.

We can make application to the stochastic neuron (e.g. Deco and Obradovic, 1996, p. 24) in which a series of inputs $y_i^j, i = 1, ...m$ from $m$ nearby neurons at time $j$ is convoluted with $m$ 'weights' $w_i^j$ at time $j$ using an inner product

$$a_j = \mathbf{y^j} \cdot \mathbf{w^j} = \sum_{i=1}^{m} y_i^j w_i^j,$$

in the context of a 'transfer function' $f(\mathbf{y^j} \cdot \mathbf{w^j})$ such that the probability of the neuron firing and having a discrete output $z^j = 1$ at time $j$ is

$$P(z^j = 1) = f(\mathbf{y^j} \cdot \mathbf{w^j}).$$

From our viewpoint the $m$ values $y_i^j$ constitute the 'sensory activity' and the $m$ weights $w_i^j$ the 'ongoing activity' at time $j$, with $a_j = \mathbf{y^j} \cdot \mathbf{w^j}$. The $w$ factors may be constant. Indeed, a systematic search for a proper set of $w$'s constitutes a 'learning paradigm' for a neural model, while the renormalization symmetry representing phase transition at 'learning plateaus' becomes an expression of neural architecture, in the large sense (R. Wallace, 2002a; Wallace et al., 2003).

Now we impose a standard thermodynamic formalism – a Legendre Transform – on the ergodic source uncertainty $H[\mathbf{X}]$ (e.g. Dembo and Zeitouni, 1998; Beck and Schlogl, 1993).

For a physical system a system's free energy density is defined as

$$F(K_1, ..., K_m) = \lim_{V \to \infty} \frac{\log[Z(K_1, ..., K_m)]}{V}$$

(2)

where $V$ is the system volume, $K_1, ..., K_m$ are other system-wide parameters, and $Z(K_1, ..., K_m)$ is the partition function defined from system's energy states. $K_1$ is usually taken as an inverse temperature.

We parametize the source uncertainty in a similar manner, so that

$$H = \lim_{n \to \infty} \frac{\log[N(n)]}{n} = H[K_1, ..., K_m, \mathbf{X}]$$

(3)

where now we take the $K_j$ as inverse 'strength of weak ties' analogs across the coupled system.

For a physical system we can obtain the equation of state which describes the macroscopic behavior of the system through imposition of a Legendre transform on equation (2). The Legendre transform of a well-behaved function $f(K_1, ..., K_m)$ is defined by

$$g = f - \sum_{i=1}^{w} K_i \partial f / \partial K_i$$

$$\equiv f - \sum_{i=1}^{w} K_i Q_i,$$

(4)

so that $Q_i = \partial f / \partial K_i$, and is invertible provided $\partial f / \partial K$ is well behaved. Then we can write

$$f = g - \sum_{i=1}^{w} Q_i \partial g/\partial Q_i.$$
(5)

The generalization when $f$ is not well-behaved is through a variational argument (Beck and Scholgl, 1993; Griffiths, 1972; Fredlin and Wentzell, 1998; Dembo and Zeitouni, 1998) rather than this tangent plane argument.

In a physical system for which $F$ is the free energy, the Legendre transform defines the macroscopic entropy as

$$S \equiv F - \sum_i K_i \partial F/\partial K_i.$$
(6)

This is the equation of state for a physical system.

We propose as a macroscopic equation of state for our coupled spatial array of stochastic resonators characterized by the source uncertainty $H[K_1,...,K_m; \mathbf{X}]$ the relation

$$S \equiv H - \sum_i K_i \partial H/\partial K_i.$$
(7)

where we will now call $S$ the macroscopic *disorder*.

We will find it necessary to significantly modify the usual treatment, which we review below.

### Large fluctuations in physical systems.

The standard discussion of large fluctuations (Onsager and Machlup, 1953; Fredlin and Wentzell, 1998) in physical systems is relevant to our analysis, as it is the principal foundation for much current study of stochastic resonance and related phenomena and thus serves as a useful reference point for our extension to cognitive systems.

The macroscopic behavior of a physical system in time is assumed to be described by the phenomenological Onsager relations giving large-scale fluxes as

$$\sum_i R_{i,j} dK_j/dt = \partial S/\partial K_i,$$
(8)

where the $R_{i,j}$ are appropriate constants.

Inverting the relations gives

$$dK_i/dt = \sum_j L_{i,j} \partial S/\partial K_j = L_i(K_1,...,K_m,t) \equiv L_i(K,t).$$
(9)

The terms $\partial S/\partial K_i$ are macroscopic driving 'forces' dependent on the entropy gradient.

Let a white Brownian 'noise' $\epsilon(t)$ perturb the system, so that

$$dK_i/dt = \sum_j L_{i,j} \partial S/\partial K_j + \epsilon(t)$$

$$= L_i(K,t) + \epsilon(t),$$
(10)

where the time averages of $\epsilon$ are $<\epsilon(t)>= 0$ and $<\epsilon(t)\epsilon(0)>= D\delta(t)$. $\delta(t)$ is the Dirac delta function, and we take $K$ as a vector in the $K_i$.

Following Luchinsky (1997), if we write the probability that the system starts at some initial macroscopic parameter state $K_0$ at time $t = 0$ and gets to the state $K(t)$ at time $t$ as $P(K,t)$, then a somewhat subtle development (e.g. Feller, 1971) gives the forward Fokker-Planck equation for $P$:

$$\partial P(K,t)/\partial t = -\nabla \cdot (L(K,t)P(K,t)) + (D/2)\nabla^2 P(K,t).$$
(11)

In the limit of weak noise intensity this can be solved using the WKB, i.e. the eikonal, approximation. Take

$$P(K,t) = z(K,t)\exp(-s(K,t)/D).$$
(12)

$z(K,t)$ is a prefactor and $s(K,t)$ is a classical action satisfying the Hamilton-Jacobi equation, which can be solved by

integrating the Hamiltonian equations of motion. The equation reexpresses $P(K,t)$ in the usual parametized negative exponential format.

Let $p \equiv \nabla s$. Substituting equation (12) in equation (11) and collecting terms of similar order in $D$ gives

$$dK/dt = p + L, dp/dt = -\partial L/\partial K p$$

$$-\partial s/\partial t \equiv h(K,p,t) = pL(K,t) + \frac{p^2}{2}$$

with $h(K,t)$ the 'Hamiltonian' for appropriate boundary conditions.

Again paraphrasing Luchinsky (1997), these 'Hamiltonian' equations have two different types of solution, depending on $p$. For $p = 0, dK/dt = L(K,t)$ which describes the system in the absence of noise. We expect that with finite noise intensity the system will give rise to a distribution about this deterministic path. Solutions for which $p \neq 0$ correspond to *optimal paths* along which the system will move with overwhelming probability.

This is a formulation of 'large fluctuation' theory which has particular attraction for physicists, few of whom can resist the magical appearance of a 'Hamiltonian.' These results can be derived, however, as a special case of a 'Large Deviation Principle' based on 'entropies' mathematically similar to Shannon's uncertainty from information theory (Dembo and Zeitouni, 1998).

Here we are concerned, not with a random Brownian distortion of simple physical systems, but with a complex 'behavioral' structure, in the largest sense, composed of quasi-independent 'actors' for which

[1] the usual Onsager relations of equations (8) and (9) may be too simple,

[2] the 'noise' may not be either small or random, and

[3] the meaningful/optimal paths have extremely high degrees of serial correlation, amounting to a grammar and syntax, precisely the fact which allows definition of the dual information source.

We will give two examples of alternative forms of 'generalized Onsager relations,' the first from a study of evolutionary process in socially-dominated populations (R. Wallace and R.G. Wallace, 1998, 1999) and the second from a relatively subtle treatment of 'resilience' in ecosystems (Ives, 1995; R. Wallace and D. Wallace, 2003b).

### Tuning the GCH: 1

The first example is similar to the usual physics treatment, with one significant change. We assume a 'social' structure to our internal self-image generated by an assembly of coupled cognitive systems that indeed responds to gradients in the disorder construct $S$, but with opposite sign to that of a physical system: A 'social' system moves away from concentrations of 'disorder' rather than towards it. Thus we have, for the $m$ parameters of $H[K_1, K_2, ...K_m, \mathbf{X}]$, $m$ equations of a 'generalized Onsager relation'

$$dK_i/dt = -L\partial S/\partial K_i, \quad (14)$$

where $L$ is positive. We adjust the system to some initial reference configuration $K_0 = K_1, K_2, ...K_m$ such that $dK/dt|_{K_0} = -L\nabla S|_{K_0} \equiv 0$.

Deviations from this reference configuration, which we write as $\delta K \equiv K - K_0$, in first order, obey the relation

$$d\delta K_i/dt \approx -L \sum_{j=1}^{m} (\partial^2 S/\partial K_i \partial K_j|_{K_0})\delta K_j. \quad (15)$$

In matrix form, writing $U_{i,j} = \partial^2 S/\partial K_i \partial K_j = U_{j,i}$, this becomes

$$d\delta K/dt = -L\mathbf{U}\delta K. \quad (16)$$

We assume some appropriate regularity conditions on $S$ and $\mathbf{U}$ and expand the deviations vector $\delta K$ in terms of the eigenvectors of the symmetric matrix $\mathbf{U}$, i.e. $m$-dimensional vectors $e_i$ such that $\mathbf{U}e_i = \lambda_i e_i$, so that

$$\delta K = \sum_{i=1}^{m} \delta a_i e_i. \quad (17)$$

Equation (16) then has the solution

$$\delta K(t) = \sum_{i=1}^{m} \delta a_i \exp(-L\lambda_i t)e_i. \quad (18)$$

Clearly any eigenconfiguration $e_j$ having a negative eigenvalue, $\lambda_j < 0$, will be amplified exponentially in time until the network goes into precisely the sudden epileptiform phase transition so extensively studied in R. Wallace (2000).

The prescription, then, is:

[1] First, identify a 'base configuration' of sensory input, a signal for which we wish to optimize the 'transmission channel' constituting our nested cognitive array. This initial step may itself involve a higher order cognitive process, since it does not seem amenable to a simple variational procedure, although measuring deviations from the base may be. This is because the base state of any cognitive physiological system is, de facto, in both a high energy and high information transmission mode, and differentiation between such states is not simply a matter of minimizing some functional, although, given such a state, deviation to first order from it may be treated as such.

[2] Since the convolution method is, presumably (although not necessarily), fixed, we must next adjust the patterns of 'ongoing activity' or highly structured 'noise' and the inner settings of the cognitive array so as to constitute a coding which maximizes the transmission rate of the channel in terms of the fixed pattern of the signal to be detected. See R. Wallace (2000) for an information theoretic discussion of this learning paradigm optimization. This will result, for fixed ongoing activity, in a reference configuration of parameters $K_0 = K_1, K_2, ...K_m$ associated with the dual information source of the generalized cognitive homunculus.

[3] Next, readjust the system until that reference state has a zero gradient in $S$: $\nabla S|_{K_0} = 0$.

[4] Specify a deviation configuration $e_j = (K_{j,1}, K_{j,2}, ...K_{j,m})$ of the parameter values $K$ which represents the desired pattern of deviation of sensory input from the chosen reference state and adjust $S$ until $e_j$ is an eigenvector of $\mathbf{U} = ||\partial^2 S/\partial K_i \partial K_j||$ having a negative eigenvalue.

[5] Thresholds can be defined by imposing a 'dimple' on $S$ near the reference state $K_0$.

Occurrence of the deviation of sensory input will, if threshold is exceeded, cause the amplification of the eigenconfiguration $e_j$ until epileptiform phase transition occurs, the 'decision rule' which defines detection.

Clearly, for a given reference configuration $K_0$, different $e_j$ with different negative eigenvalues will lead to slightly different forms of epileptiform detection transitions.

Continuing the theoretical development another step, we let $d\delta K_i/dt \equiv \delta V_i$, we can rewrite our first order approximation so as to obtain an expression for the magnitude of $(\delta V)^2$:

$$(\delta V)^2 \approx L^2/2 \sum_{i,j} [\sum_k U_{i,k} U_{k,j}] \delta K_i \delta K_j.$$

(19)

Defining

$$g_{i,j} \equiv L^2/2 \sum_k (\partial^2 S/\partial K_i \partial K_k)(\partial^2 S/\partial K_k \partial K_j)$$

(20)

produces something much like the fundamental relation of a Riemannian differential geometry:

$$dV^2 = \sum_{i,j} g_{i,j}(K) dK_i dK_j.$$

(21)

A geodesic represents, in this configuration, not a minimization of $V$ along some path in $K$-space, but rather its maximization, equivalent to a minimization of the time-of-flight. The model is of a ball bearing rolling down a hill. A quasi-stochastic extension would see a Brownian fuzz around optimal paths. Not unexpectedly, we have recovered something much like the results of the previous section: The next treatment, however, is considerably more devious.

### Tuning the GCH: 2

Our second approach to the question is by means of Ives' (1995) theory of resilience for complex ecological systems, which are almost always plagued by multiple feedback loops. This does not appear to simply reduce to the standard physics model with a negative sign.

The usual vision of Onsager relations assumes that rates of change of the defining parameters of a physical system are in direct proportion to gradients in the entropy. Above we argued that broadly 'social' systems would be driven by the *negative* of such a gradient. Here we will assume that complex 'behavioral' systems are indeed affected by the entropy-analog we have called the 'disorder,' but in a very general way, seeking feedbacks in the equation of state of the information system, defined by empirical measures of biomarkers, beliefs, feelings, and behaviors, having the form

$$S(t) = H[(S(t))] - \sum_i K_i(S(t)) \partial H/\partial K.$$

(22)

Like the conventional Onsager relations we start with a first order effect, near some reference configuration $(S_0, K_0)$. Writing deviations in $S$ and the $K_i$ from that configuration on the same footing as variates $x_j$, we will seek empirical 'Onsager-like' regression relations in the deviations rather than in $\partial S/\partial K_i$ and $dK_i/dt$:

$$x_j(t) = \sum_{k \neq j}^m b_{j,k} x_k(t) + b_{j,0} + \epsilon(t, x_1(t), ...x_m(t)).$$

The $\epsilon$ terms include both 'noise' and nonlinearities, and are not necessarily small.

In matrix notation,

$$X(t) = \mathbf{B}X(t) + U(t),$$

(24)

where $\mathbf{B} = \mathbf{B}|_{S_0, K_0}$ is a fixed $m \times m$ matrix of regression coefficients having a zero diagonal and $U(t)$ is an $m$-dimensional vector containing both the constant and 'error' terms.

'Error' terms are taken as including 'shocks' outside the internal feedback loops. We are thus assuming that the coupled spatial array of stochastic resonators is operating normally in terms of the convoluted input stream of 'sensory' and 'ongoing' activity, and we ask about the effect of perturbations from this 'normal' state $(S_0, K_0)$.

We begin by rewriting the matrix equation as

$$[\mathbf{I} - \mathbf{B}]X(t) = U(t)$$

(25)

where $\mathbf{I}$ is the $m \times m$ identity matrix and, to reiterate, $\mathbf{B}$ has a zero diagonal.

We now reexpress matters *in terms of the eigenstructure of* $\mathbf{B}$.

Let $\mathbf{Q}$ be the matrix of eigenvectors which diagonalizes $\mathbf{B}$. Take $\mathbf{Q}Y(t) = X(t)$ and $\mathbf{Q}W(t) = U(t)$. Let $\mathbf{J}$ be the diagonal matrix of eigenvalues of $\mathbf{B}$, so that $\mathbf{B} = \mathbf{Q}\mathbf{J}\mathbf{Q}^{-1}$. In R. Wallace et al. (1997) we show how, after some nontrivial development, the treatment can be reduced to canonical correlation, so that the eigenvalues of $\mathbf{B}$ are all real. Then, for the eigenvectors $Y_k$ of $\mathbf{B}$, corresponding to the eigenvalues $\lambda_k$,

$$Y_k(t) = \mathbf{J}Y_k(t) + W_k(t).$$

(26)

Using a term-by-term shorthand for the components of the non-orthogonal $Y_k$, this becomes

$$y_k(t) = \lambda_k y_k(t) + w_k(t).$$

(27)

In the usual manner define the mean of a time-dependent function $f(t)$ over the interval $\Delta T$ as

$$E[f(t)] \equiv \frac{1}{\Delta T} \int_0^{\Delta T} f(t)dt,$$

(28)

and the variance as $Var[f] = E[(f - E[f])^2]$. We assume an appropriately rational structure as $\Delta T \to \infty$.

Again taking matters term-by-term, get the variance in the $y_k$ as $Var[(1-\lambda_k)y_k] = Var[w_k]$, so that

$$Var[y_k] = \frac{Var[w_k]}{(1-\lambda_k)^2},$$

or, taking $\sigma = \sqrt{Var}$,

$$\sigma(y_k) = \frac{\sigma(w_k)}{|1-\lambda_k|}.$$

(29)

.

The $y_k$ are the components of the eigentransformed deviation variates $x_i$, and the $w_k$ are the similarly transformed variates of the driving externalities $u_i(t)$.

The eigenvectors $Y_k$ are characteristic but non-orthogonal combinations of the original variates $x_i$ whose standard deviation is that of the particular patterns of driving externalities $W_k$, but synergistically amplified by the term $1/|1-\lambda_k|$, a function of the eigenvalues of the matrix of regression coefficients $\mathbf{B}$. This kind of amplification is typical in ecosystem theoretics (e.g. Caswell, 1999), but seems difficult to obtain using the standard eikonal treatment.

For a two dimensional system it is easy to show that $\lambda$ is just the correlation between the two variates.

We are suggesting, then, that a system consisting of sensory activity, ongoing activity and a nested array of cognitive processes can be finely tuned so that the internal parameters of the system will strongly *and characteristically* respond, in particular, to small deviations of the parameters associated with the 'sensory activity' from a given pattern, i.e. to characteristic patterns of 'noise' $W_k$. A systematic tuning process involving error minimization would constitute a 'learning paradigm' in the sense of neural networks.

The non-orthogonal nature of the excited eigenstates $Y_k$, however, means that this response will leak into other characteristic tuned eigenmodes, resulting in a retina-analog within the GCH:

Suppose we rewrite a GCH response to short-term perturbation – not the effects of long-lasting structured psychosocial stress – as

$$X_{t+1} = (\mathbf{B}_0 + \delta\mathbf{B}_{t+1})X_t,$$

where $X_t$ is the state at time $t$ and $\delta\mathbf{B}_{t+1}$ now represents the effect of perturbation.

Again we impose a (Jordan block) diagonalization in terms of the matrix of (generally nonorthogonal) eigenvectors $\mathbf{Q}_0$ of the 'zero reference configuration' $\mathbf{B}_0$, obtaining, for an initial condition which is an eigenvector $Y_t \equiv Y_k$ of $\mathbf{B}_0$,

$$Y_{t+1} = (\mathbf{J}_0 + \delta\mathbf{J}_{t+1})Y_k = \lambda_k Y_k + \delta Y_{t+1} =$$

$$\lambda_k Y_k + \sum_{j=1}^n a_j Y_j,$$

(30)

where $\mathbf{J}_0$ is a (block) diagonal matrix as above, $\delta\mathbf{J}_{t+1} \equiv \mathbf{Q}_0 \delta\mathbf{B}_{t+1}\mathbf{Q}_0^{-1}$, and $\delta Y_{t+1}$ *has been expanded in terms of a spectrum of the eigenvectors of* $\mathbf{B}_0$, with

$$|a_j| \ll |\lambda_k|, |a_{j+1}| \ll |a_j|.$$

(31)

The essential point is that, provided $\mathbf{B}_0$ has been properly tuned, so that this condition is true, the first few terms in the spectrum of the plieotropic iteration of the eigenstate will contain almost all of the essential information about the perturbation, i.e. most of the variance. We envision this as similar to the detection of color in the optical retina, where three overlapping non-orthogonal 'eigenmodes' of response suffice to characterize a vast array of color sensations. Here, if a concise spectral expansion is possible, a very small number of (typically nonorthogonal) 'generalized cognitive eigenmodes' permit characterization of a vast range of external perturbations, and rate distortion constraints become very manageable indeed. Thus GCH responses – the spectrum of excited eigenmodes of $\mathbf{B}_0$, provided it is properly tuned – can be a very accurate and precise gauge of environmental perturbation.

The choice of zero reference state $\mathbf{B}_0$, i.e. the 'base state' from which perturbations are measured, is, we claim, a highly nontrivial task, necessitating a specialized, most likely higher-order cognitive, apparatus. Again, this is because all states of a cognitive physiological system are, relatively speaking, high energy, high information, states, so that no simple variational procedure can identify a base state, although it may measure deviations from such a state.

### Extending the theory: disjointly, locally and nearly ergodic information sources

Following the treatment of Cover and Thomas, (1991, p. 474), the Shannon-McMillan Theorem on which we have based our analysis is predicated on having a stationary ergodic information source – one whose long-time pattern of emitted symbols follows the strong law of large numbers. An ergodic source is defined on some probability space $(\Omega, \mathcal{B}, \mu)$, where $\mathcal{B}$ is a sigma algebra of subsets of the space $\Omega$ and $\mu$ is a probability measure. A random variable $X$ is defined as a function $X(\omega), \omega \in \Omega$, on the probability space. There is also a time translation operator, $T: \Omega \to \Omega$. Let $\mu$ be the probability measure of a set $A \in \mathcal{B}$. Then the transformation is *stationary* if $\mu(TA) = \mu(A)$ for all $A \in \Omega$. The transformation is *ergodic* if every set $A$ such that $TA = A$ almost everywhere satisfies $\mu(A) = 0$ or 1. That is, almost everything flows.

If the transformation $T$ is stationary and ergodic, we call the process defined by $X_n(\omega) = X(T^n\omega)$ stationary and ergodic.

For a stationary ergodic source with a finite expected value, the Ergodic Theorem concludes that

$$\frac{1}{n}\sum_{i=1}^n X_i(\omega) \to E(X) = \int X d\mu$$

with probability 1. This is the generalized law of large numbers for ergodic processes: the arithmetic mean in time converges to the mathematical expectation in 'space.'

Beginning here, after some considerable mathematical travail, the Shannon-McMillan Theorem, as we have described it, follows (Khinchine, 1957; Petersen, 1995; Cover and Thomas, 1991).

The essential point is that for a stationary, ergodic information source the limit

$$H[\mathbf{X}] = \lim_{n\to\infty} \frac{H[X_0,...X_n]}{n+1}$$

not only exists, but *is independent of path*. That is, as $x = a_0,...,a_n$ gets longer and longer, all paths converge to the same value of $H[\mathbf{X}]$ regardless of their origin or meandering. This is the fundamental information theory simplification, onto which we have imposed parametization and on which we have further grafted invariance under renormalization at 'learning plateau' phase transition as an expression of architecture.

A careful reading of the proof to the Shannon-McMillan Theorem (Khinchine, 1957; Petersen, 1995) shows that non-ergodic information sources still converge to some value $\lim_{n\to\infty} H(x)$, where $x$ is a path of increasing length, but the value $H(x)$ is now *path dependent*. That is, each increasing path $x$ converges to its own value of $H(x)$, depending, thus, on both the overall 'language' and on the particular path chosen.

If the underlying state space can be partitioned into disjoint equivalence classes of meaningful paths, then we may break the system into mutually disjoint information sources, and proceed as above, much like working with separate domains of attraction in a nonlinear system. We will call such a system 'disjointly' ergodic. This structure would allow an organism to shift between different languages of thought to meet different classes of externalities.

A second way of proceeding is 'locally,' i.e. imposing a 'manifold' structure on the underlying state space, the collection of paths $x$. That is, we assume a topology for the state space such that each path $x$ has an open neighborhood which can be mapped by a homoeomorphism onto a reference state space which has an ergodic information source. With appropriate topology, each open covering of the state space has a finite subcovering which patches the thing together in the standard manner (e.g. Sternberg, 1964; Thirring, 1992). This differential geometry approach is recognizably similar to, but would seem to generalize, use of an 'information metric' to derive asymptotic statistical results (Amari, 1982; Kass, 1989).

An explicit attack in this spirit is to suppose that, given one particular highly probable path, $x_0$, we can reasonably define the source uncertainty associated with nearby paths in terms of their distance from $x_0$. Let $x = x_0 + \delta x$, where $\delta x$ represents a 'small variation,' and make the usual formal series expansion near $H(x_0) \equiv H_0$ in terms of a generalized derivative:

$$H(x) = H_0 + \delta H(x) \approx H_0 + \frac{\delta H}{\delta x} \delta x,$$

where we assume $\delta H \ll H_0$.

We might well call such a system 'nearly' ergodic.

Extension of our treatment to slightly-less-than ergodic systems appears possible, showing that cognitive processes may indeed exist beyond language-of-thought models strongly constrained by grammar and syntax. All three approaches would seem amenable to a higher order tuning in which the organism 'shifts gears' to meet changing demands.

### Discussion and conclusions.

We find it plausible that cognitive processes having dual information sources also have equations of state, and invoke various generalized Onsager relations to describe the macroscopic response of the system to the entropy-analog we have termed the disorder. Subsequent development leads to possibly large amplification effects within a generalized cognitive self-image, the GCH, which might be tuned – much in the sense of the learning paradigms of neural networks – to detect deviation from expected reference patterns of normality, i.e. a tunable retina. This suggests that a retina-analog of the GCH may be particularly and characteristically efficient at detecting deviations from the base state of an organism, including its social relations. Designation of such a base state, however, since there is no 'natural' functional whose extremal can be harnessed for its identification, would seem to be a matter of a secondary, embedding, cognitive or developmental process. Extremal formalism could, to reiterate, be useful for detecting changes from that baseline in the GCH.

We have imposed retina dynamics in the form of Onsager relations in terms of a large deviations analog and a linear regression model. Other approaches are clearly possible. Thus we suggest that, in addition to internal language structure, renormalization symmetry representing underlying architecture, in the largest sense, and state-space partitioning, cognitive physiological or psychosocial modules might have to be classified as well by their particular Onsager relations, which determine response to the disorder variate arising naturally from a Legendre transform applied to their parametized source uncertainty, once the base-state has been identified by other means.

Onsager relations for cognitive or other information systems, however, might not at all resemble the simple linear response in rate of change of parameters to gradients in the entropy which characterizes physical systems. R. Wallace (2000) similarly suggested that the renormalization relations representing underlying architecture, in the large sense, which are associated with information arrays might bear little or no resemblance to familiar physical pictures, a matter explored at length in R. Wallace and R.G. Wallace (2003) or Wallace et al. (2003).

The payoff for introducing these complications is an indication of the 'natural' direction for characterizing the behavior of cognitive systems which do not have a simple duality with an ergodic information source, i.e. for which even generalized language-of-thought arguments fail. We are, ultimately, suggesting that languages-of-thought may be amenable to some form of 'second order selection' analogous to that described by the mutator formalism of Wallace et al. (2003) and R. Wallace and R.G. Wallace (2003). Such systems, which could adjust their grammar and syntax according to the short-term demands facing the organism, are likely to have exceedingly rich response repertoires, providing considerable long-term adaptive advantage under evolutionary selection pressures.